\documentclass[12pt,preprint]{aastex}

\shorttitle{Entropy at the Outskirts}
\shortauthors{Fujita et al.}

\begin{document}


\title{Entropy at the Outskirts of Galaxy Clusters as Implications
for Cosmological Cosmic-Ray Acceleration}

\author{Yutaka Fujita}
\affil{Department of Earth and Space Science, Graduate School of
 Science, Osaka University, Toyonaka, Osaka 560-0043, Japan}
\email{fujita@vega.ess.sci.osaka-u.ac.jp}

\and

\author{Yutaka Ohira and Ryo Yamazaki}
\affil{Department of
Physics and Mathematics, Aoyama Gakuin University, Fuchinobe, Chuou-ku,
Sagamihara 252-5258, Japan}

\begin{abstract}
 Recently, gas entropy at the outskirts of galaxy clusters has attracted
 much attention. We propose that the entropy profiles could be used to
 study cosmic-ray (CR) acceleration around the clusters. If the CRs are
 effectively accelerated at the formation of clusters, the kinetic
 energy of infalling gas is consumed by the acceleration and the gas
 entropy should decrease. As a result, the entropy profiles become flat
 at the outskirts. If the acceleration is not efficient, the entropy
 should continue to increase outwards. By comparing model predictions
 with X-ray observations with {\it Suzaku}, which show flat entropy
 profiles, we find that the CRs have carried $\lesssim 7$\% of the
 kinetic energy of the gas away from the clusters. Moreover, the CR
 pressure at the outskirts can be $\lesssim 40$\% of the total
 pressure. On the other hand, if the entropy profiles are not flat at
 the outskirts as indicated by combined {\it Plank} and {\it ROSAT}
 observations, the carried energy and the CR pressure should be much
 smaller than the above estimations.
\end{abstract}


\section{Introduction}

The existence of cosmic-rays (CRs) in clusters of galaxies is clearly
indicated by radio synchrotron emission \citep{fer12a}. The CRs may be
accelerated at shocks or turbulence in the X-ray hot gas
\citep*[e.g.][]{fuj01,bru01,fuj03a,kan05,pfr07a}, and they may play
important roles in the evolution of clusters. For example, they may heat
the cores of clusters \citep{fuj11b,fuj12a,fuj13a}. Since the background
gas is hot, their acceleration efficiency in clusters could be very
different from that of objects in the Galaxy (e.g. supernova remnants;
SNRs). However, the lack of firm observational evidence of CRs with the
exception of radio observations prevent us from understanding their
acceleration. Thus, other observational approaches that can clarify
CR acceleration in clusters are strongly desired.

X-ray observations with {\it Suzaku} have revealed the entropy of the
hot gas of clusters at their outskirts. Here, the entropy is defined as
$K\equiv k_B T/n^{\gamma_g-1}$, where $k_B$ is the Boltzmann constant,
$\gamma_g$ ($=5/3$) is the adiabatic index for the gas, and $T$ and $n$
are the gas temperature and the number density, respectively. Numerical
simulations, in which non-gravitational heating (e.g. energy ejection
from galaxies) and radiative cooling are not included, predicted that
the entropy should increase with $r$ as $\propto r^{1.1}$, where $r$ is
the distance from the cluster center \citep*{voi05a,bur10a}. On the
other hand, the {\it Suzaku} observations have shown that the entropy
profiles flatten at $r\gtrsim 0.5 \: r_{200}$
\citep{bau09,kaw10,hos10,aka11,sim11a,sat12a,wal12a,ich13a} (see
Figure~13 of \citealt{sat12a}), where $r_{200}$ ($\sim 2$~Mpc) is the
radius within which the mean mass density is 200 times the critical
density for a flat universe and is often considered to be the typical
radius of clusters.

Several ideas have been proposed to solve this discrepancy. Gas clumping
could lead to the overestimation of the gas density, which in turn could
lead to the underestimation of entropy \citep{nag11a}. However, gas
clumping is effective mainly at $r\gtrsim r_{200}$, because the clumps
tend to be destroyed within the cluster ($r\lesssim r_{200}$).
Non-equilibrium between ions and electrons may also decrease
entropy. This is because heavier ions have most of the kinetic energy of
gas and they are first thermally heated at shocks. If it takes a long
time for the kinetic energy of the ions to be transferred to lighter
electrons, the observed temperature of the electrons could be low. This
may decrease the observed entropy of gas \citep{hos10,aka11}. However,
numerical simulations have shown that the degree of non-equilibrium is
small \citep{won09a}, although it may depend on the dynamical state of
clusters \citep{aka08,rud09a}. Moreover, from observations of the
Sunyaev-Zel'dovich effect, {\it Planck} has shown that the gas pressure
at the outskirts of clusters does not significantly drop. This suggests
that the ions and electrons are in equilibrium \citep{pla13a}. For some
clusters, accretion of matter toward them may decrease at low
redshifts. Although this makes the entropy profiles flatter at their
outskirts, it does not seem to be effective enough to be consistent with
the observed systematic flattering \citep{cav11a}.

In this Letter, we propose that the entropy profiles may reflect CR
acceleration in clusters and could be used as a tool to study it. We
focus on the acceleration at 'accretion shocks' around the
clusters. These shocks are generated through gas infall toward the
clusters as they grow by accreting dark matter and gas from the outside
\citep*[e.g.][]{evr96,ryu03,ski08a}. If the CR acceleration at the
shocks is effective, the CRs should consume much of the kinetic energy
of the gas and should affect the entropy of the gas \citep{byk05}. Thus,
the flattened entropy profiles may be the result of the past CR
acceleration around the clusters.

\section{Models}

It is widely believed that if CRs are effectively accelerated at a
shock, their pressure changes the structure of the shock
\citep{dru81,dru83,mal01}. Before we investigate the entropy profiles of
clusters, we consider the jump conditions at the shock. We treat gas and
CRs as two fluids hereafter. We define three distinct regions with a
shock: (0) the far upstream region where the influence of accelerated
CRs can be ignored, (1) a region just upstream of the shock where the
presence of the CRs affects the shock structure, and (2) a region
downstream from the shock. We indicate each region by lower indices. For
example, the gas density and velocity in the far upstream region are
$\rho_0$ and $v_0$, respectively. The spatial scale of the region (1) is
much smaller than the size of a cluster, and it cannot be resolved by
current X-ray telescopes. The gas density and temperature could
respectively change from $\rho_2$ and $T_2$ far downstream of the shock
as the cluster grows.

The CR acceleration consumes gas energy. We discuss the accompanied
entropy change at a shock based on a model of \citet{vin10a}. We
represent the energy flux that is carried away by CRs diffusing away far
upstream by $F_{\rm cr}$. The flux is normalized to the kinetic energy
flux of the shock:
\begin{equation}
 \epsilon_{\rm esc} \equiv \frac{F_{\rm cr}}{(1/2)\rho_0 v_0^3}\:.
\end{equation}
The fraction of the downstream CR pressure is
\begin{equation}
 w \equiv \frac{P_{\rm 2,cr}}{P_2} = \frac{P_{\rm 2,cr}}{P_{\rm 2,th} +
  P_{\rm 2,cr}}
\end{equation}
where 'cr' and 'th' represent the CR and the thermal components,
respectively. The Mach number of the shock defined at the region far
upstream is
\begin{equation}
 {\cal M}_0 \equiv \sqrt{\frac{1}{\gamma_g}\frac{\rho_0 v_0^2}{P_0}} \:.
\end{equation}
The compression ratios across the different regions are
\begin{equation}
 \chi_1 \equiv \frac{\rho_1}{\rho_0},\; 
 \chi_2 \equiv \frac{\rho_2}{\rho_1},\;
 \chi_{12} \equiv \chi_1 \chi_2 = \frac{\rho_2}{\rho_0}\:.
\end{equation}
Assuming that the thermal component evolves adiabatically in the region
(1), the pressure is given by $P_{\rm 1,th}=P_0
\chi_1^{\gamma_g}$. Thus, the total compression ratio is given by
\begin{equation}
\label{eq:chi12}
 \chi_{12}=\frac{(\gamma_g + 1){\cal M}_0^2
  \chi_1^{-\gamma_g}}
{(\gamma_g-1){\cal M}_0^2\chi_1^{-(\gamma_g+1)}+2}\:.
\end{equation}
The CR pressure fraction is given by
\begin{equation}
 w = \frac{(1-\chi_1^{\gamma_g}) + \gamma_g {\cal M}_0^2(1-1/\chi_1)}
{1 + \gamma_g {\cal M}_0^2(1-1/\chi_{12})}\:, 
\end{equation}
and the escaping energy flux is given by
\begin{equation}
 \epsilon_{\rm esc} = 1+\frac{2G_0}{\gamma_g {\cal M}_0^2}
-\frac{2G_2}{\gamma_g {\cal M}_0^2 \chi_{\rm 12}}
-\frac{2G_2}{\chi_{\rm 12}}
\left(1-\frac{1}{\chi_{12}}\right)-\frac{1}{\chi_{12}^2}\:,
\end{equation}
where
\begin{equation}
\label{eq:G}
 G_0 \equiv \frac{\gamma_g}{\gamma_g-1},\; 
G_2 \equiv w\frac{\gamma_{\rm cr}}{\gamma_{\rm cr}-1}
+ (1-w)\frac{\gamma_g}{\gamma_g-1}\;,
\end{equation}
and $\gamma_{\rm cr}$ is the adiabatic index for the CR component.
Equations.~(\ref{eq:chi12})--(\ref{eq:G}) show that $\epsilon_{\rm esc}$
and $w$ are represented by $\chi_1$ for a given ${\cal M}_0$. The
entropy difference across the shock is written as
\begin{equation}
\label{eq:DS}
 \Delta S = \frac{3}{2}k_B\ln
\left(\frac{K_2}{K_0}\right)\:.
\end{equation}
In this equation,
\begin{equation}
\label{eq:KK}
 \frac{K_2}{K_0} = (1-w)\frac{1}{\chi_{12}^{\gamma_g}}
\left[1 + \gamma_g {\cal M}_0^2 
\left(1 - \frac{1}{\chi_{12}}\right)\right]\:,
\end{equation}
which means that $\Delta S$ is also the function of ${\cal M}_0$ and
$\chi_1$ \citep{vin10a}.

\begin{figure}
\epsscale{.80}
\plotone{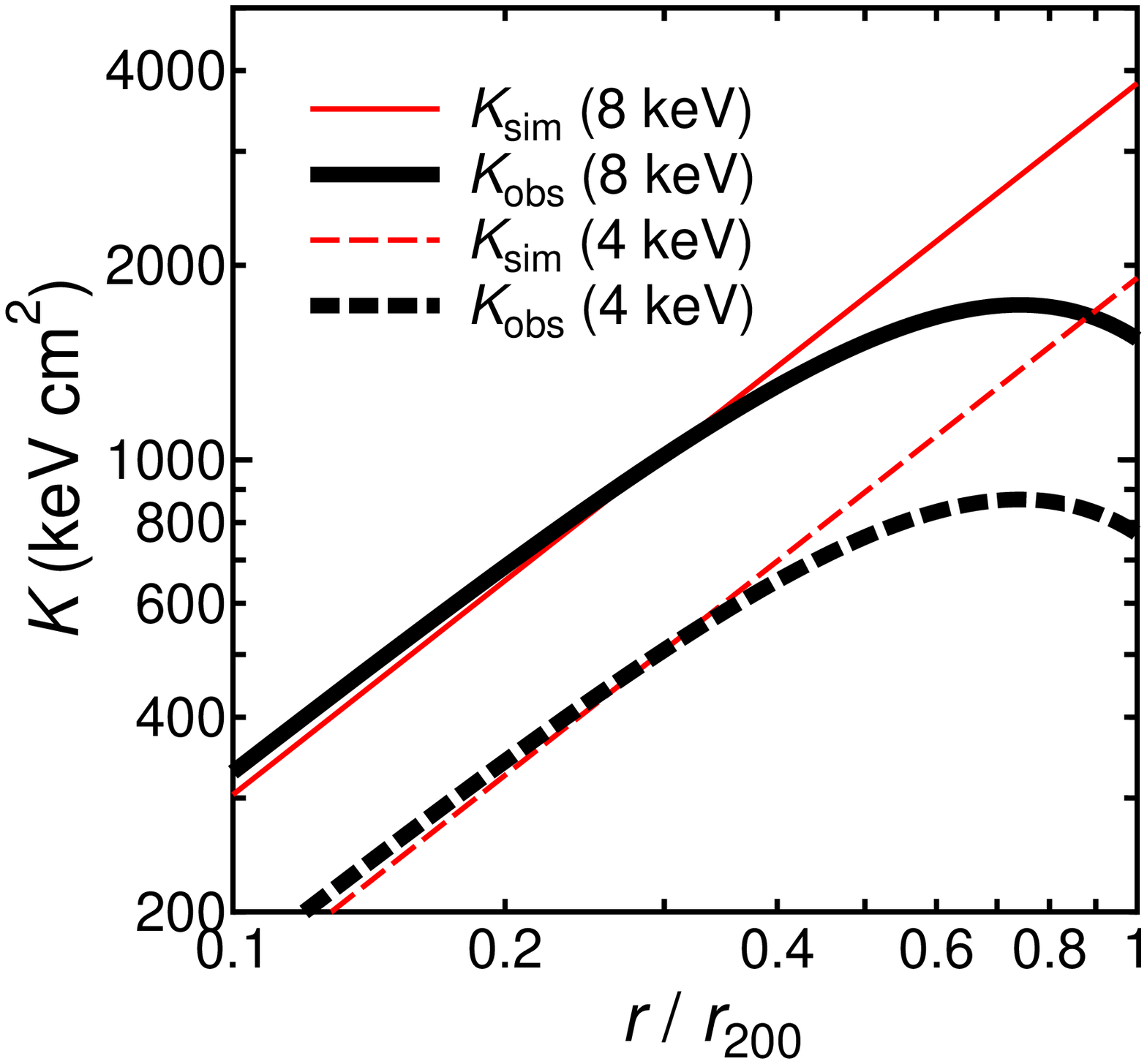}
 \caption{Profiles of $K_{\rm sim}$ (thin lines) and $K_{\rm obs}$
 (thick lines) for
 $T_X=8$~keV (solid lines) and $T_X=4$~keV (dashed lines)} 
\label{fig:K}
\end{figure}

It would be natural to assume that CRs are accelerated at the accretion
shocks \citep{fuj01,kan05,pfr07a}, although clear evidence has not been
discovered. The CRs would carry the gas energy away and lower the
entropy. We compare the entropy profiles predicted by numerical
simulations that do not include CR acceleration with those obtained
through X-ray observations. Contrary to density and temperature, entropy
is conserved if there is no heating and cooling, which is an advantage
to discuss entropy. The entropy profile predicted by the numerical
simulations is
\begin{equation}
\label{eq:Ksim}
 K_{\rm sim}(r)=1.32\: K_{200}(r/r_{200})^{1.1}\:,
\end{equation}
where $K_{200}\approx 362\;{\rm keV\; cm^2}(T_X/{\rm keV})$ and $T_X$ is
the gas temperature \citep{voi05a}.  If we adopt the entropy profiles
obtained with {\it Suzaku}, they can be fitted as
\begin{equation}
\label{eq:Kobs}
 K_{\rm obs}(r) \propto (r/r_{200})^{1.1}e^{-(r/r_{200})^2}
\end{equation}
\citep{wal12b}. At $r\sim 0.3\: r_{200}$, the entropy profiles do not
become flat and they are not much affected by the activities of central
active galactic nuclei (AGNs). Thus, we assume that
\begin{equation}
\label{eq:Kobsn}
 K_{\rm obs}(0.3 r_{\rm 200}) = K_{\rm sim}(0.3 r_{\rm 200}) \:,
\end{equation}
which gives the normalization of Equation~(\ref{eq:Kobs}). Although
there is a debate about an appropriate position of the normalization
($r\sim 0.3\: r_{200}$; \citealt{eck13a}), it would not affect the
following results when the observed entropy profiles significantly
deviate from those predicted by the theoretical
simulations. Figure~\ref{fig:K} shows $K_{\rm sim}(r)$ and $K_{\rm
obs}(r)$ for $T_X=8$ and 4~keV. Although $K_{\rm obs}$ slightly
decreases at $r\gtrsim 0.7\: r_{200}$, the curve should be regarded as
'flat' for $r\gtrsim 0.5\: r_{200}$ considering the errors of the
observations. We note that while non-thermal pressure owing to
turbulence in gas may affect the entropy profile \citep*{lau09a}, it
should be included in $K_{\rm sim}$ obtained by high-resolution
numerical simulations that can follow turbulence. Thus, the difference
between $K_{\rm sim}$ and $K_{\rm obs}$ is made by something other than
the non-thermal pressure.

We assume that a cluster gradually grows from the inside to the outside
as it accretes matter from the outside. The entropy of the gas that
falls into the cluster may not be zero, because it could have been
heated through activities of galaxies and AGNs in the field
\citep*{kai91,cav97,pon99}. In fact, widely distributed metals
outside clusters may indicate such activities \citep{fuj08}. The entropy
of the gas before passing accretion shocks is estimated to be $K_{\rm
out}\sim 100$--$400\:\rm keV\: cm^2$ \citep{voi03}, and we adopt $K_{\rm
out}=K_0=200\:\rm keV\: cm^2$.

We ignore heating and cooling of the gas except for the heating at
accretion shocks around the cluster ($r\gtrsim r_{200}$). The motion of
the gas is fundamentally controlled by the gravity from dark matter, and
thus the infall velocity of the gas toward the cluster and the Mach
number of the accretion shocks (${\cal M}_0$) do not significantly
depend on the presence or absence of CR acceleration at the shocks,
while the postshock gas (region 2) is significantly affected by
it. Therefore, the Mach number of the shock ${\cal M}_0={\cal M}(r)$,
through which the gas located at the radius $r$ at present had passed,
can be estimated using $K_{\rm sim}(r)$ and $K_{\rm out}$. Since $K_{\rm
sim}(r)$ is the entropy when the CR acceleration is ignored, the
relation among ${\cal M}$, $K_{\rm sim}$ and $K_{\rm out}$ should be
described by the Rankine-Hugoniot relation that does not include the
effects of CRs ($\chi_1=1$):
\begin{equation}
\label{eq:RH}
 \frac{K_{\rm sim}(r)}{K_{\rm out}}=\frac{2\gamma_g{\cal M}(r)^2 
- (\gamma_g - 1)}{\gamma_g  + 1}
\left[\frac{(\gamma_g - 1){\cal M}(r)^2 + 2}
{(\gamma_g + 1){\cal M}(r)^2}\right]^{\gamma_g}\:.
\end{equation}
Figure~\ref{fig:M} shows the profiles of the Mach number ${\cal M}(r)$
derived from Equation~(\ref{eq:RH}). The Mach number is the increasing
function of $r$ and reaches ${\cal M}\sim 10$ at $r\sim r_{200}$.  We
emphasize that entropy is conserved for a given gas element, and that
the gas observed at $r$ at present was not necessarily there when it
passed a shock. Therefore, {\it the radius of the shock could be larger
than $r_{200}$ at the time when the gas passed the shock.} In other
words, the position of the shock is not specified in our model.

Assuming that CR acceleration makes $K_{\rm obs}$ smaller than $K_{\rm
sim}$ at the outskirts of clusters, the entropy difference at the shocks
for the observed clusters is obtained by using Equations~(\ref{eq:DS})
and~(\ref{eq:KK}) with ${\cal M}_0={\cal M}(r)$:
\begin{equation}
\label{eq:DS2}
\Delta S[{\cal M}(r), \chi_1(r)] = \frac{3}{2}k_B\ln
\left[\frac{K_{\rm obs}(r)}{K_{\rm out}}\right] \:.
\end{equation}
Since we already know $K_{\rm obs}(r)$ (Equations~[\ref{eq:Kobs}]
and~[\ref{eq:Kobsn}]) and ${\cal M}(r)$ (Figure~\ref{fig:M}), we can
derive $\chi_1(r)$ from Equation~(\ref{eq:DS2}).

\section{Results and Discussion}

From ${\cal M}_0={\cal M}(r)$ and $\chi_1(r)$, we obtain $\epsilon_{\rm
esc}(r)=\epsilon_{\rm esc}[{\cal M}(r), \chi_1(r)]$ and $w(r)=w[{\cal
M}(r), \chi_1(r)]$ using Equations~(\ref{eq:chi12})--(\ref{eq:G}), which
are shown in Figure~\ref{fig:ew}. In the calculations, we assumed that
$\gamma_{\rm cr}=4/3$. Figure~\ref{fig:ew} shows that $\epsilon_{\rm
esc}$ and $w$ increase outwards and reach $\epsilon_{\rm esc}\sim 0.07$
and $w\sim 0.4$ at $r\sim r_{200}$. These values are often estimated for
the Galactic SNRs \citep{vin10a}, although their Mach numbers are
typically an order of magnitude larger than those of the cluster shocks
considered in this study. Nevertheless it is interesting that clusters
have similar values of $\epsilon_{\rm esc}$ and $w$. Since other factors
such as clumping of the gas \citep{nag11a} or a decrease of matter
accretion toward the clusters \citep*{cav11a} may partially contribute
to the decrease of the entropy, the actual values of $\epsilon_{\rm
esc}$ and $w$ could be somewhat smaller than the above values.

\begin{figure}
\epsscale{.80}
\plotone{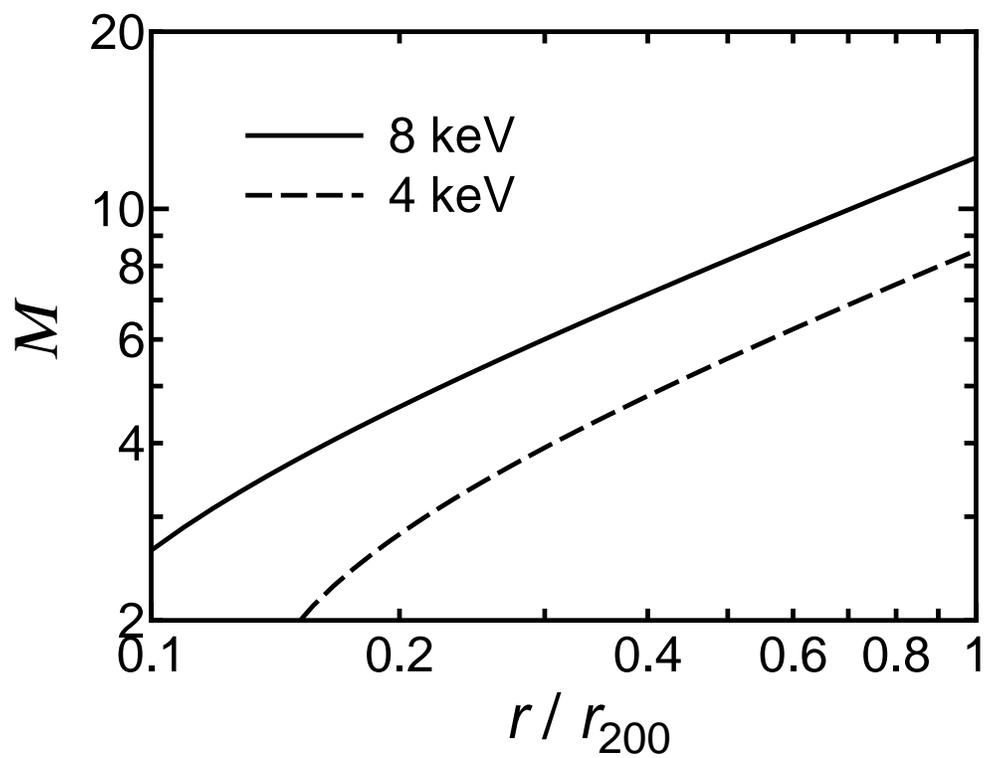}
 \caption{Profiles of ${\cal M}$ for
 $T_X=8$~keV (solid line) and $T_X=4$~keV (dashed line).} 
\label{fig:M}
\end{figure}

\begin{figure}
\epsscale{.80}
\plotone{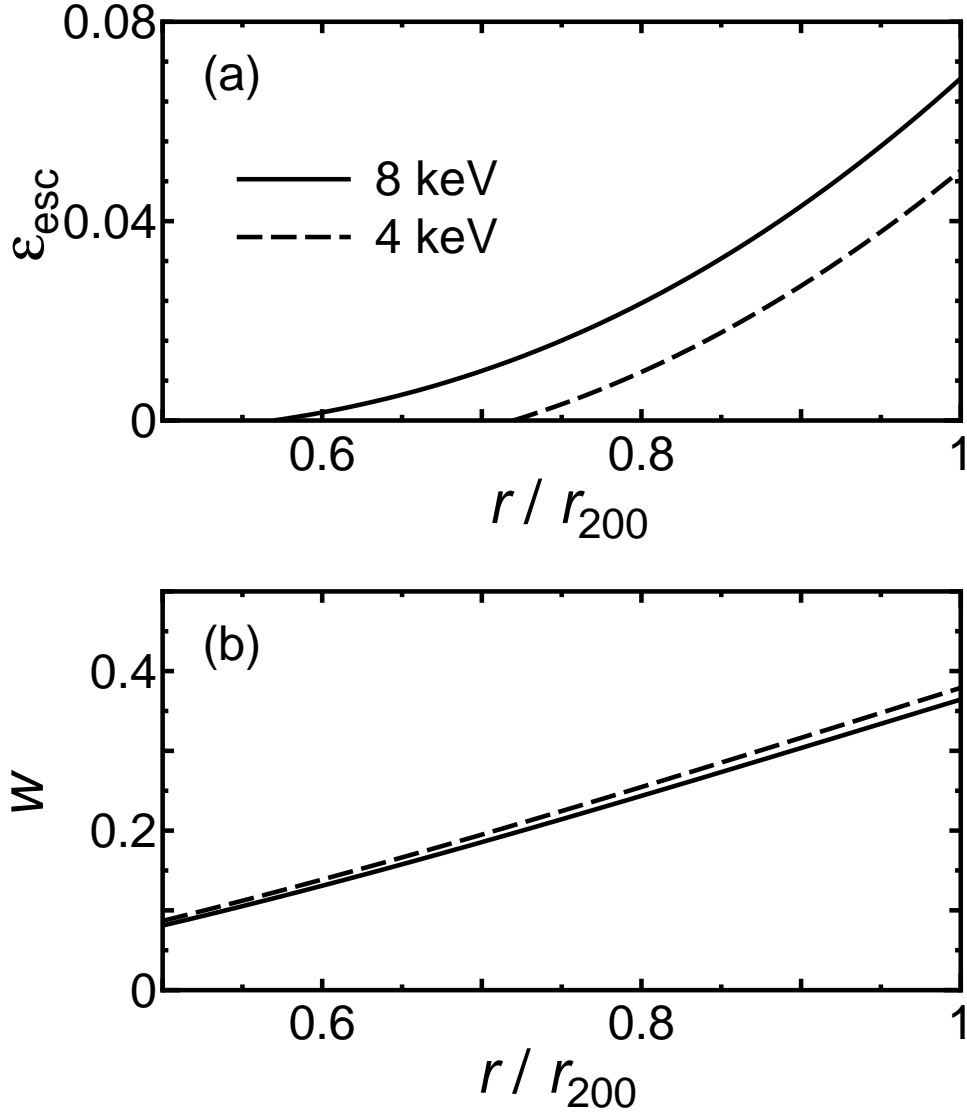}
 \caption{Profiles of (a) $\epsilon_{\rm esc}$ and (b) $w$ for
 $T_X=8$~keV (solid lines) and $T_X=4$~keV (dashed lines).} 
\label{fig:ew}
\end{figure}

We implicitly assumed that the accretion shock is a single
shock. However, the accreted gas may have passed multiple shocks. In
that case, the Mach number of the inner shocks may be smaller than
${\cal M}$ in Figure~\ref{fig:M}, because of the heating at the outer
shocks. Thus, our model could not be applied to dynamically active
clusters that have complicated shock structures. However, the CRs may be
reaccelerated at multiple shocks and the acceleration efficiency may be
large in spite of the low Mach numbers \citep{kan11a}. Therefore, CR
acceleration at multiple shocks with small Mach numbers may be
qualitatively similar to that at a shock with a large Mach number.  At
least we can say that the difference between $K_{\rm obs}$ and $K_{\rm
sim}$ indicate that a significant fraction of the gas energy must have
been converted to CRs even in the case of multiple shocks.

Some cosmological numerical simulations do not show the drastic
temperature decrease or density increase at cluster outskirts in spite
of CR acceleration \citep[e.g.][]{pfr07a,vaz12a}. Although we could not
identify the cause, it may be because their assumed acceleration
efficiency is smaller than those we obtained. In fact, if a constant
high efficiency of 50\% is adopted, the temperature significantly drops
\citep[Figure~3 of][]{pfr07a}. Moreover, the efficiency in the central
region of a cluster may need to be much lower than that at the outskirt
in order to reproduce the flat entropy profile. In our model, the CR
acceleration is prohibited in the central region because of the low Mach
number attributed to the preheated gas ($K_{\rm out}>0$). This leads to
a low density of CR protons and less interaction of CR protons with gas
protons in the central region. This may be favorable for the
non-detection of gamma-rays from the central regions of clusters
\citep{ack10a}, because gamma-rays are produced through proton-proton
interaction.

Our model predicts that entropy decreases in the region where CR
acceleration is effective. The CR pressure is $w\gtrsim 0.1$, which is
the typical value for the Galactic SNRs \citep{vin10a}, at $0.5\:
r_{200}\lesssim r \lesssim r_{200}$ (Figure~\ref{fig:ew}b). In this
region, synchrotron emission from electrons ('radio relics') has often
been discovered \citep{fer12a}. Although the electrons seem to be
accelerated at shocks created during cluster mergers, the Mach numbers
are relatively small (${\cal M}\sim 2$--4) \citep{van11b,aka13a}. At
these small Mach numbers, CR acceleration may be difficult. However, if
there is a pre-existing CR population, CR reacceleration at the shocks
could increase their energy high enough to emit the synchrotron
radiation \citep{kan11a}. The entropy decrease at cluster outskirts may
indicate the existence of a pre-existing CR population. Moreover, the
CRs may amplify magnetic fields
\citep*{luc00a,bel04a,fuj10a,fuj11a}. Thus, magnetic fields amplified by
the CRs may be found at the cluster outskirts in the future. The CRs
accelerated at the accretion shocks may contain ultra-high-energy ones
\citep*{ino03}.

Note that the flat entropy profiles have not been confirmed in the
combined analysis of the pressure profiles obtained with {\it Plank} and
the density profiles obtained with {\it ROSAT} \citep{eck13a}. In that
study, the entropy continues to increase outward. If this is the case,
our model indicates that the CR acceleration efficiency is much smaller
than those estimated above. This could constrain the aforementioned
reacceleration model for the radio relics. This also shows that the
future confirmation of the entropy profiles is important in terms of CR
acceleration in clusters. At present, the model and observational
uncertainties suggest that the estimated values of $\epsilon_{\rm
esc}\sim 0.07$ and $w\sim 0.4$ at $r\sim r_{200}$ at the beginning of
this section should be regarded as upper limits.

\acknowledgments

We appreciate the referee's useful comments. We thank F.~Takahara and
D.~Nagai for useful discussions. This work was supported by KAKENHI (YF:
23540308, YO: No.24.8344).

\clearpage

\clearpage

\clearpage

\end{document}